\documentclass[
    ,final            % use final for the camera ready runs
  ,numberedheadings % uncomment this option for numbered sections
  ]
  {aipproc}

\layoutstyle{6x9}

\usepackage{epsfig,multicol}
\usepackage{amssymb}
\usepackage{graphics}
\usepackage{amsmath,latexsym}
\def\beq{\begin{eqnarray}}    %%%  begequation/eqnarray
\def\eeq{\end{eqnarray}}      %%%  endequation/eqnarray

\newcommand{\Om}{\Omega_m}
\newcommand{\Omo}{\Omega_m^0}

\newcommand{\OMo}{\Omega_{M}^0}
\newcommand{\ORo}{\Omega_{R}^0}
\newcommand{\OL}{\Omega_{\Lambda}}
\newcommand{\Oc}{\Omega_{c}}

\newcommand{\OLo}{\Omega_{\Lambda}^0}
\newcommand{\OX}{\Omega_{X}}
\newcommand{\OXo}{\Omega_{X}^0}

\newcommand{\OD}{\Omega_{D}}
\newcommand{\ODo}{\Omega_{D}^0}

\newcommand{\OKo}{\Omega_{K}^0}

\newcommand{\rc}{\rho_c}
\newcommand{\rco}{\rho_{c}^0}

\newcommand{\rmr}{\rho_m}

\newcommand{\rD}{\rho_D}

\newcommand{\rX}{\rho_X}
\newcommand{\pX}{p_X}
\newcommand{\wX}{\omega_X}
\newcommand{\wm}{\omega_m}
\newcommand{\wR}{\omega_R}

\newcommand{\amr}{\alpha_m}
\newcommand{\aef}{\alpha_e}
\newcommand{\aX}{\alpha_X}
\newcommand{\rL}{\rho_{\CC}}
\newcommand{\pL}{p_{\CC}}

\newcommand{\pD}{p_D}

\newcommand{\CC}{\Lambda}

\newcommand{\we}{\omega_{e}}

\newcommand{\lu}{\lambda_1}
\newcommand{\ld}{\lambda_2}

\begin{document}

\title{$\Lambda$XCDM cosmologies: solving the cosmological coincidence problem?}


\author{Javier Grande}{
  address={High Energy Physics Group, Dep. ECM,  Univ. de
Barcelona,\\ \hspace{0.2cm}
Diagonal 647, 08028 Barcelona, Catalonia, Spain\\
\hspace{0.2cm} E-mails: jgrande@ecm.ub.es, sola@ifae.es, stefancic@ecm.ub.es\\}
}

\author{Joan Sol\`a$\;\!$\small{$^{1}$}\normalsize{$^\textrm{,}$}}{
  address={High Energy Physics Group, Dep. ECM,  Univ. de
Barcelona,\\ \hspace{0.2cm}
Diagonal 647, 08028 Barcelona, Catalonia, Spain\\
\hspace{0.2cm} E-mails: jgrande@ecm.ub.es, sola@ifae.es, stefancic@ecm.ub.es\\}
  ,altaddress={C.E.R. for Astrophysics, Particle Physics and
Cosmology\,\footnotemark[2]\footnotetext[2]{Associated with Instituto de Ciencias del
Espacio-CSIC.}}
}

\author{Hrvoje
\v{S}tefan\v{c}i\'{c}$\;\!$} %\small{$^{2}$}\normalsize{$^\textrm{,}$}}
{
  address={High Energy Physics Group, Dep. ECM,  Univ. de
Barcelona,\\ \hspace{0.2cm}
Diagonal 647, 08028 Barcelona, Catalonia, Spain\\
\hspace{0.2cm} E-mails: jgrande@ecm.ub.es, sola@ifae.es, stefancic@ecm.ub.es\\} % additional visiting address
,altaddress={Theoretical Physics Division, Rudjer Bo\v{s}kovi\'{c}
Institute, P.O.B. 180, HR-10002 Zagreb, Croatia.}
}
\footnotetext[1]{Speaker. Invited talk at DSU 2006, Madrid, June 20-24 2006.}
%\footnotetext[2]{On leave of absence from
%the Theoretical Physics Division, Rudjer Bo\v{s}kovi\'{c}
%Institute, Zagreb, Croatia.}
\setcounter{footnote}{2}

\begin{abstract}
We explore the possibility of having a composite (self-conserved)
dark energy (DE) whose dynamics is controlled by the quantum
running of the cosmological parameters. We find that within this
scenario it is feasible to find an explanation for the
cosmological coincidence problem and at the same time a good
qualitative description of the present data.
\end{abstract}

\maketitle

%%%%%%%%%%%%%%%%%%%%%%%%%%%%%%%%%%%%%%%%%%%%
%% MAINMATTER
%%%%%%%%%%%%%%%%%%%%%%%%%%%%%%%%%%%%%%%%%%%%

\section{Introduction}

Independent data from different observations\,\cite{Supernovae,
WMAP3Y,LSS} provide strong support for the existence of DE and
seem to agree that it presently constitutes $\sim70\%$ of the
total energy density. Nevertheless, the nature of DE remains
unclear. If we identify it with a cosmological constant (CC)
arising from the quantum field theory (QFT) vacuum energy, as
done in the $\CC$CDM model\,\cite{Peebles84}, we are led to a
value many orders of magnitude greater than the measured one,
what has been called the CC problem \cite{weinRMP,Copeland06}.
This problem could be alleviated by means of a dynamical DE. This
possibility is supported by some recent
studies\,\cite{Alam,Jassal1} and has been exploited profusely in
various forms \cite{Copeland06}. Among them the scalar fields are
the most paradigmatic scenario. It must be stressed though that
the presence of a scalar field is not essential for a model to be
described in terms of an effective EOS, $\pD=\we\,\rD$ (for
instance, this has been proven for any model with variable
cosmological parameters in \cite{SS12}).

We present here a model in which the DE, in addition of being
dynamical, is allowed to be composite. This model may offer an
explanation to the ``cosmological coincidence
problem''\,\cite{weinRMP} -i.e. to the fact that the DE and matter
densities happen to be similar precisely at the present epoch- by
keeping the ratio between these two densities bounded and of
order 1 during the entire Universe existence. At the same time,
the effective EOS of our model can match the available data. This
feature is exemplified through specific numerical examples.

\section{The $\Lambda$XCDM model}\label{sec:pm}

The $\CC$XCDM \cite{GSS1} is a minimal realization of a composite
DE model in which the DE consists of a running $\CC$\,\cite{nova}
and another entity, $X$, that may interact with it. We call this
new entity the ``cosmon''\,\cite{PSW}. The nature of $X$ is not
specified, although a most popular possibility would be some
scalar field $\chi$ resulting e.g. from string theory at low
energy. In the generalized sense defined here, the cosmon stands
for any dynamical contribution to the DE other than the vacuum
energy effects.

The model under study contains matter-radiation
($\rho_m=\rho_M+\rho_R$) and dynamical DE
$\left(\rho_D(t)=\rX(t)+\rL(t)\right)$, where the two DE
components have barotropic indices $\wX$ and
$\omega_{\Lambda}=-1$. We also suppose that  $X$ can be both
quintessence (QE) ($-1<\wX<-1/3$) or phantom-like ($\wX<-1$), that
is: $\ -1-\delta<\wX<-1/3\ (\delta>0)$. Assuming $G=const.$
(another realization of the $\CC$XCDM model in which G can also be
variable is considered in \cite{GSS2}) and the conservation of
matter-radiation, the Bianchy identities give us:
\begin{equation}
\begin{array}{rrl}
&\dot{\rho}_m+\amr\,\rmr\,H=0\,,&\ \ \ \ \amr\equiv3(1+\wm)\,.\\
\dot{\rho}_D+\alpha_e\,\rD\,H=0
\longrightarrow&\dot{\rho}_{\Lambda}+\dot{\rho}_X+\,\aX\,\rX\,H=0\,,&\
\ \ \ \aX\equiv 3(1+\wX)\,.
\end{array}\label{conslawDE2}
\end{equation}
where $\wm=0,1/3\ (\amr=3,4)$ for the matter and the radiation
dominated epoch respectively. The effective EOS parameter of the
model reads:
\begin{equation}\label{eEOS}
\we=\frac{\pD}{\rD}=\frac{\pL+\pX}{\rL+\rX}=\frac{-\rL+\wX\,\rX}{\rL+\rX}=
-1+(1+\wX)\,\frac{\rX}{\rD}\,.
\end{equation}
Another fundamental equation for our model is Friedmann's
equation:
\begin{equation}
H^{2}\equiv \left( \frac{\dot{a}}{a}\right) ^{2}=\frac{8\pi\,G }{3}%
\left( \rmr +\rD\right) -\frac{k}{a^{2}}=\frac{8\pi\,G }{3}%
\left( \rmr +\rL+\rX\right) -\frac{k}{a^{2}}\,.  \label{FL}
\end{equation}
From it, the generalized form of the `cosmic sum rule' within our
model is easily derived:
\begin{equation}\label{sumrule0}
\OMo+\ODo+\OKo=\OMo+\OLo+\OXo+\OKo=1\,,\ \ \ \
\Omega_i^0\equiv\frac{\rho_i^0\phantom{^0}}{\rho_c^0}=\frac{8\pi
G\rho_i^0}{3H_0^2},\ \ \ \ \ \Omega_K^0\equiv\frac{-k}{H_0^2}\,.
\end{equation}
We still need another equation apart from (\ref{conslawDE2}),
(\ref{FL}), so we have to provide a model either for $X$ or for
$\rL$. We will do the latter in order to preserve the generic
nature of $X$, its dynamics being then determined by the Bianchi
identity through (\ref{conslawDE2}). Following \,\cite{RGTypeIa},
we adopt the following RG equation:
\begin{equation}\label{RGEG1b}
\frac{d\rL}{d\ln \mu}=\frac{3\,\nu}{4\,\pi}\,M_P^2\,\mu^2\,,
\end{equation}
where $\mu$ is the energy scale associated to the RG in Cosmology
(that we will identify with the Hubble parameter at any given
epoch,\,\cite{RGTypeIa}) and $\nu$ is a free parameter that
essentially provides the squared ratio of the heavy masses
contributing to the $\beta$-function of $\CC$ versus the Planck
mass, $M_P$ (and thus we naturally expect $\nu\ll 1$). Equation
(\ref{RGEG1b}) with $\mu=H(t)$ is the equation we were looking
for in order to solve the model.

But before that, let us take a closer look at the implications of
having a composite DE. Taking the derivative of (\ref{FL}) and
using as well (\ref{conslawDE2}) we obtain:
\begin{equation}
\frac{\ddot{a}}{a}=-\frac
{4\pi\,G}{3}\,[\rmr\,\amr+\rD\,\aef]=-\frac
{4\pi\,G}{3}\,[\rmr\,(1+3\wm)-2\,\rL+\rX\,(1+3\wX)]\,.\label{dda}
\end{equation}
Note that if $\wX <-1/3$ \emph{but also} $\rho_X<0$, then the $X$
component decelerates the expansion instead of accelerating it.
And this is perfectly possible in our model thanks to the
composite nature of the DE: e.g., at present (\ref{sumrule0})
tells us -in the flat case- that $\ODo=\OXo+\OLo=1-\Omo>0$ but
either $\OXo$ or $\OLo$ could be negative. In particular, it can
occur that $\rho_X<0$ and $\omega_X<-1$, the cosmon being
therefore phantom-like \emph{but} with $p_X>-\rho_X>0$  (in
contrast to the ''standard'' phantom condition $p_i<-\rho_i<0$).
Therefore, $X$ fulfills the strong energy condition (SEC)
-satisfied by matter but violated by ``usual'' phantom and QE
components-, behaving thus like a sort of unclustered ``matter''
that we call ``Phantom matter'' (PM), see Fig.\ref{fig1}(a). This
behavior is possible in any model with composite DE.

We will see (c.f. Sect.\ref{sec:coin}) that the solution to the
coincidence problem is linked to the existence of a point $z_s$
where the Universe expansion stops (and subsequently reverses),
i.e. $H(z_s)=0$. Although in our model we can have $\CC_0<0$,
this stopping point can be achieved even if $\CC_0>0$ thanks to
the behavior of $X$ as PM.

\section{Solution of the $\CC$XCDM model}

From now on we will assume spatial flatness and constant $\wX$
(cf. \cite{GSS1} for the general case). Instead of $t$ or $z$ we
will use $\zeta=-\ln (1+z)$ as the independent variable $\left(t=0
\leftrightarrow \zeta=-\infty\right.$, $t=t_0 \leftrightarrow
\zeta=0$, $\left.t=\infty \leftrightarrow \zeta=\infty\right)$.
In this way our basic set of equations becomes an autonomous
system:
\begin{eqnarray}\label{autonomous1}
&&\dot{\Omega}_X=-\left[\nu\,\amr+(1-\nu)\,\aX\right]\,\OX-\nu\,\amr\,\OL+\nu\,\amr\,\Oc\,,\nonumber\\
&&\dot{\Omega}_{\CC}=\nu\,(\amr-\aX)\,\OX+\nu\,\amr\,\OL-\nu\,\amr\,\Oc\,,\nonumber\\
&&\,\dot{\Omega}_c=(\amr-\aX)\,\OX+\,\amr\,\OL-\amr\,\Oc\,,
\end{eqnarray}
where $\dot{\phantom{A}} \equiv d/d\zeta$ and
$\Oc(z)=\rc(z)/\rco=H^2(z)/H_0^2$. Here all $\Omega_i(z)$ are
normalized to the present critical density, $\rho_c^0$. The
solution of the system reads:
\begin{equation}\label{solve2}
\bar{\Omega}(\zeta)\equiv\begin{pmatrix} \OX(\zeta) \\ \OL(\zeta)
\\ \Oc(\zeta) \end{pmatrix}\equiv\big(\OX,\OL,\Oc\big)^t=
C_1\,{\bf v_1}\,e^{\lambda_1\,\zeta}+C_2\,{\bf
v_2}\,e^{\lambda_2\,\zeta}+C_3\,{\bf v_3}\,,
\end{equation}
with:
\begin{equation}
{\setlength\arraycolsep{7pt}
\begin{array}{lll}
\lambda_1=-\aX\,(1-\nu)\,,&\lambda_2=-\amr\,,&\lambda_3=0\,.\\[1ex]
{\bf v_1}=\big(1-\nu,\nu,1\big)^t,&{\bf v_2}=\Big(\frac{-\nu\,\amr}{\amr-\aX},\nu,1\Big)^t\,,&{\bf v_3}=\big(0,1,1\big)^t\,.\\[1.5ex]
C_1=1-C_2-C_3\,,&C_2=\frac{\Omo(\amr-\aX)}{\amr-\aX\,(1-\nu)}\,,&C_3=\frac{\OLo-\nu}{1-\nu}\,,\\
\end{array}}\label{eig}
\end{equation}
where the constants $C_j$ result from the boundary conditions at
present: $\Omega_i(\zeta=0)=\Omega_i^0$.

\subsection{Nucleosynthesis bounds and the coincidence problem}\label{sec:coin}
The expansion rate is sensitive to the amount of DE, and
therefore primordial nucleosynthesis can place stringent bounds
on the parameters of the $\CC$XCDM model. We will ask for the
ratio between DE and matter radiation densities to be
sufficiently small at the nucleosynthesis epoch,
$|r(z=z_N\sim10^9)|\lesssim10\%$ (\cite{GSS1,RGTypeIa}, see also
\cite{Ferreira97}). From (\ref{solve2}):
\begin{equation}\label{rz}
r(z)=\frac{\rD}{\rmr}=\frac{C_3+C_1\,(1+z)^{\aX\,(1-\nu)}+(C_2-\Omo)\,(1+z)^{\amr}}{\Omo(1+z)^{\amr}}\,,
\end{equation}
where we have returned to $z$ as the independent variable for a
while. At $z=z_N$ we can neglect $C_3$ in the numerator and
(remembering we are in the radiation era) we get:
\begin{equation}\label{rzN}
r_N\equiv
r(z_N)=-\frac{\epsilon}{\wR-\wX+\epsilon}+\frac{C_1}{\ORo}\,(1+z_N)^{-3\,(\wR-\wX+\epsilon)},\
\ \ \ \ \ \epsilon\equiv\nu\,(1+\wX)\,.
\end{equation}
Now, keeping in mind that $-1-\delta<\wX<-1/3$ and that $\wR=1/3$,
it is easy to see that:
\begin{equation}\label{nb1}
|r_N|<10\%\,\Longleftrightarrow\,\frac{|\epsilon|}{\wR-\wX+\epsilon}\simeq|\epsilon|=|\nu\,(1+\wX)|<0.1\,.
\end{equation}
Note that for $\nu\neq 0$ there is an irreducible contribution of
the DE to the total energy density in the radiation era. Looking
again at (\ref{rz}), but this time at the dark energy dominated era, we
find that $r(z)$ can present (at most) one extremum at some
$z=z_e$\,\cite{GSS1}. Let us prove that the existence of a future
stopping of the expansion (feature that can occur within our
model, c.f. Sect.\ref{sec:num}) implies that of a future maximum
of $r(z)$ -and viceversa-. By (\ref{FL}) and (\ref{dda}):
{\setlength\arraycolsep{1pt}\begin{eqnarray}
&&\lim_{z\rightarrow-1}H^2/H_0^2=\lim_{z\rightarrow-1}\OD\,,\label{FL-1}\\
&&\left.\frac{\ddot{a}}{a}\right|_{t=t_0}=-\frac{4\pi\,G}{3}\,[(1+r_0)+r'(0)]\rho_m(0)
\,,\label{acc}
\end{eqnarray}}
\noindent where $r_0\equiv r(z=0)$ and $r'(z)=dr(z)/dz$. Note
that, since $r_0>0$, the current state of accelerated expansion
requires  $r'(0)<0$. Now, if the RHS of (\ref{FL-1}) is positive,
$\lim_{z\rightarrow-1}r(z)=\infty$ and the ratio is unbounded.
Moreover, as $r'(0)<0$ and the $r(z)$ is a smooth function with
\emph{at most} one extremum\,\cite{GSS1}, there cannot be any
extremum in the future, and thus the DE can't get negative and
there is no stopping point. On the contrary, if the RHS of
(\ref{FL-1}) is negative, as $H(z)$ is also continuous, then
$H(z_s)=0$ at some $z_s>-1$. Being $r_0>0$, $r'(0)<0$, and
$\lim_{z\rightarrow -1}r(z)=-\infty$ it is obvious that there
must be a maximum of $r(z)$ at some point between $z=0$ and
$z=z_s\ $ (\textit{q.e.d.}).

\subsection{Behavior of the EOS in the far past: a signature of the model}
From the solution of the model (\ref{solve2}), we find that in
the asymptotic past and for $\nu\neq0$:
\begin{equation}
{\setlength\arraycolsep{1pt}
\begin{array}{rlr}
\OD(z\gg 1)=&-\frac{\epsilon}{\wm-\wX+\epsilon}\,\Omo\,(1+z)^{\amr}\,,&(\nu\neq0)\\
\we(z\gg 1)=&-1+(1+\wX)\,\frac{\OX(z\gg 1)}{\OD(z\gg 1)}=\wm\,, \
\ \ \ &(\nu\neq 0)\,.
\end{array}}
\end{equation}
This comes as a bit of a surprise: at very high redshift the
effective EOS of the DE coincides with that of matter-radiation.
This behavior could be detected given that it enforces:
\begin{equation}\label{renormalization}
H^2(z\gg1)\simeq H_0^2\,\hat{\Omega}_m^0\,(1+z)^{\amr}\,, \ \ \ \
\ \
\hat{\Omega}_m^0=\Omo\,\left(1-\frac{\epsilon}{\wm-\wX+\epsilon}\right)\,.
\end{equation}
That means that the measures of the parameter $\Omo$ from CMB
fits (high $z$) and supernovae data fits (low $z$) could differ,
the relative difference $|\hat{\Omega}_m^0-\Omo|/\Omo$ being just
given by the nucleosynthesis constraint (\ref{nb1}). Thus the
effect could amount to a measurable $10\%$, what makes it a
distinctive signature of the $\CC$XCDM model.

\section{Numerical analysis of the model}\label{sec:num}

Let us illustrate our considerations with some examples. Taking
the prior $\OMo=0.3\,$\,\cite{LSS}, we are left with three free
parameters: $(\OLo,\wX,\nu)$, over which we will impose that:
\begin{itemize}
\item i) The nucleosynthesis bound (the \emph{exact} one in (\ref{nb1}))
is fulfilled: $|r_N|<10\%\,$;
\item ii) There is a stopping point in the future Universe evolution;
\item iii) The ratio $r(z)$ is not only bounded
(what is guaranteed by the stopping of the expansion)
 but also stays relatively small, say $r(z)<10\cdot r_0\,\ \forall\,z\in (-1,\infty)\,$.
\end{itemize}

\begin{figure}
\mbox{\resizebox*{0.5\textwidth}{!}{\includegraphics{figure1.eps}}\
\ \ \
   \resizebox*{0.45\textwidth}{!}{\includegraphics{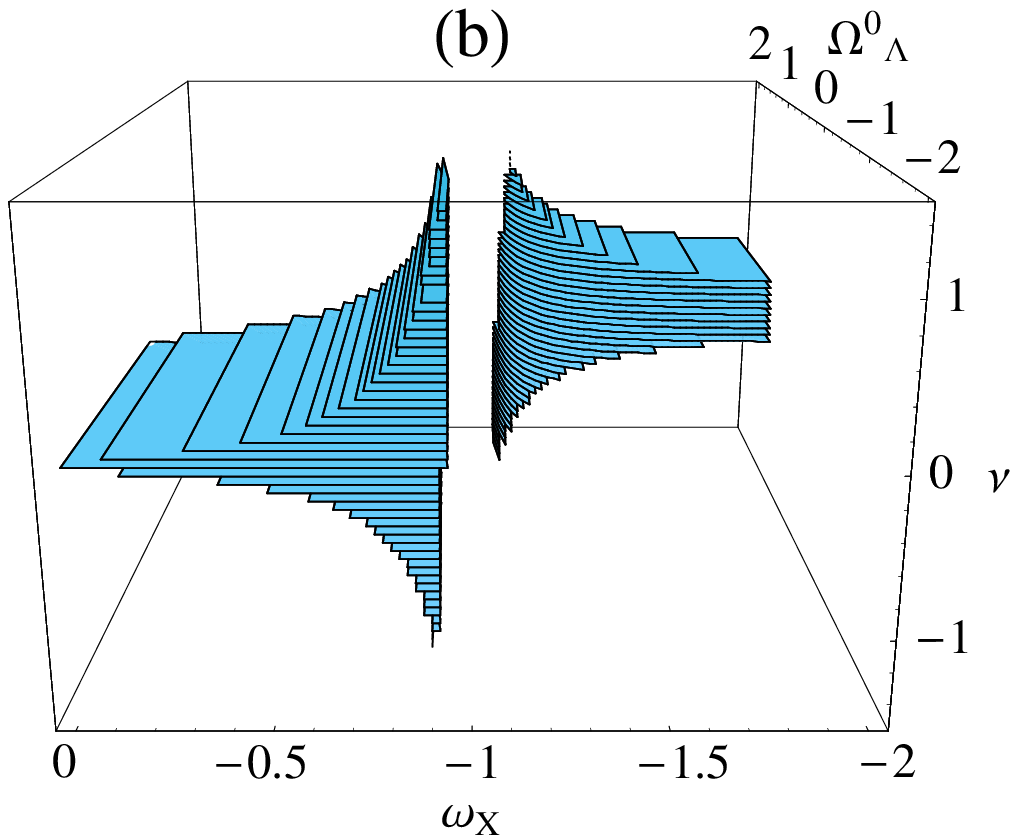}}}
\caption{(a) The Weak (WEC, $\rho\ge0$ and $\rho+p\ge0$),
Dominant (DEC, $\rho\ge|p|$ ) and Strong (SEC, $\rho+p\ge0$ and
$\rho+3p\ge0$) energy conditions. The ``standard'' QE and phantom
regions as well as the ``Phantom Matter'' ($\we<-1$ with
$\rho<0$) one are also shown; (b) The $3D$ volume formed by the
points of the $\CC$XCDM parameter space that satisfy: i) the
nucleosynthesis bound (\ref{nb1}), ii) there is a turning point
in the Universe evolution and iii) the relation
$r(z)=\rD/\rmr<10r_0$ holds for the whole Universe lifetime\,.}
\label{fig1}
\end{figure}

The points satisfying all three conditions constitute a
significant part of the full parameter space as seen in the $3D$
plot in Fig.\ref{fig1}(b). As an example, we consider the
specific situation $\ -1-\delta<\wX<0$ and $\nu<1$. Looking at the
system (\ref{autonomous1}), we see that $\lu>0\,,\ld<0$, so there
is a saddle point in the phase space,
$\bar{\Omega}^*=\left(0,\OLo,\OLo\right)^t$, from which
trajectories diverge with the evolution (as
$\zeta\rightarrow\infty$). This runaway, however, can be stopped
provided $C_1<0$ in (\ref{eig}). Indeed, since the eigenvector
${\bf v_1}$ defines a runaway direction, if $C_1<0$ the third
component of (\ref{solve2}) will eventually become negative, and
there will be a stopping. Using (\ref{nb1}), the stopping
condition acquires the form:
\begin{equation}\label{C1approx}
C_1=1-C_2-C_3=\frac{1-\OLo}{1-\nu}-\frac{\OMo(\wm-\wX)}
{\wm-\wX+\epsilon}\simeq\frac{1-\OLo}{1-\nu}-\OMo<0\,.
\end{equation}

\begin{figure}
\mbox{\resizebox*{0.45\textwidth}{!}{\includegraphics{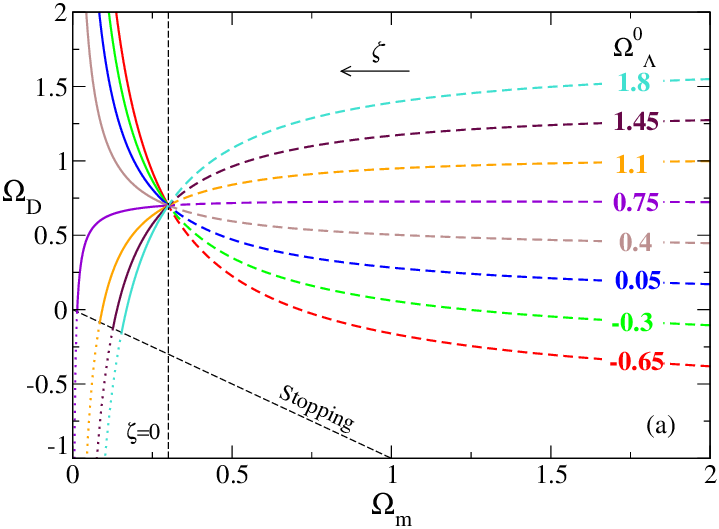}}\
\ \ \
   \resizebox*{0.45\textwidth}{!}{\includegraphics{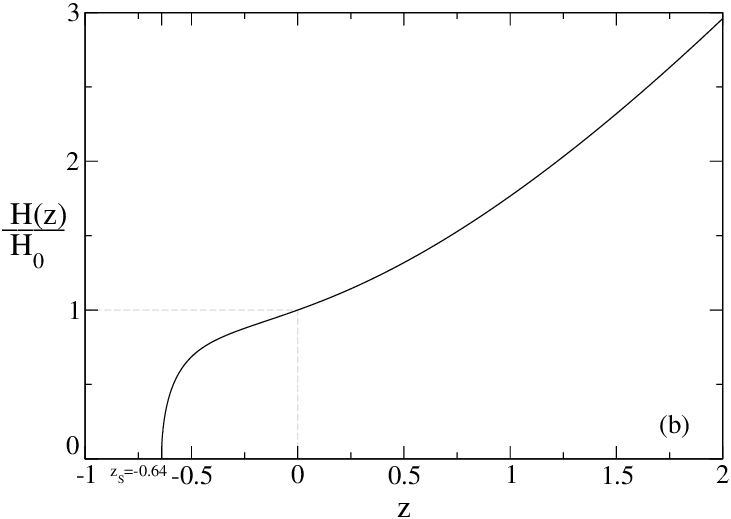}}}
\caption{(a) Phase trajectories of the autonomous system
(\ref{autonomous1}) in the $(\Om,\OD)$ plane corresponding to
$\wX=-1.85$, $\nu=-\nu_0$ ($\nu_0\equiv 0.026$) and differents
choices of $\OLo$. Dashed lines show the parts of the curves
corresponding to our past, full lines the parts between the
present moment and the stopping (if there is stopping) and dotted
lines the inaccessible part of the trajectory after the stopping;
(b) the Hubble function $H=H(z)$ for the curve with: $\OLo=0.75$,
showing the existence of a turning point.} \label{fig2}
\end{figure}

These features can be seen in Fig.\ref{fig2}(a), where the
trajectories corresponding to a fixed value  of $\wX$ and $\nu$
and various values of $\OL$ have been plotted in the ($\Om,\OD$)
plane. Only the curves that fulfil (\ref{C1approx}) get stopped.
The Hubble function of one of the stopped trajectories is plotted
in Fig.\ref{fig2}(b), showing indeed the existence of a turning
point, that in this case, as discussed in Sect.\ref{sec:pm}, is
due to the behavior of the cosmon as PM.

\begin{figure}
\mbox{\resizebox*{0.45\textwidth}{!}{\includegraphics{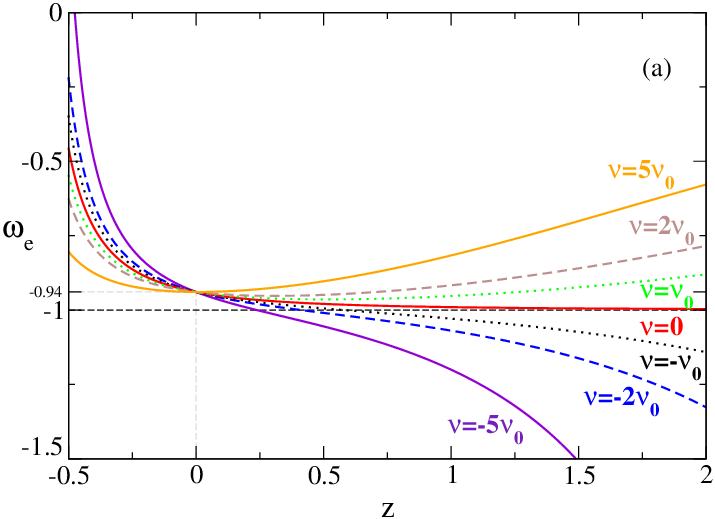}}\
\ \ \
   \resizebox*{0.45\textwidth}{!}{\includegraphics{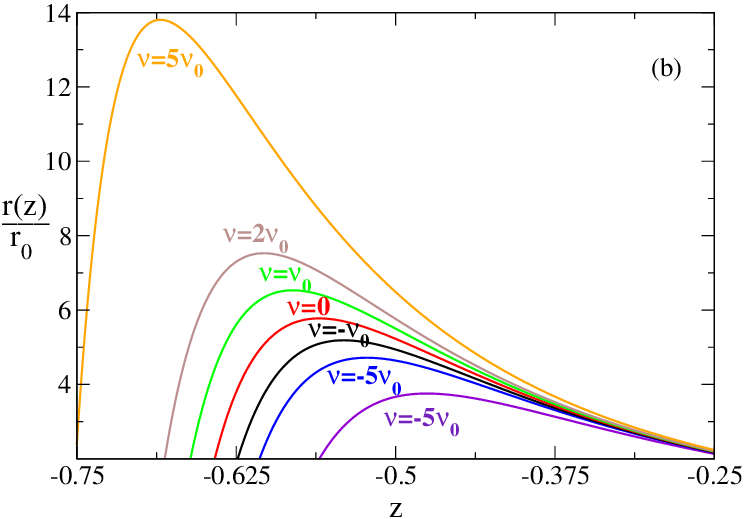}}}
\caption{ (a) Comparison of the effective EOS parameter of the
$\CC$XCDM model, $\we$, for $\wX=-1.85$, $\OLo=0.75$ and
different values of $\nu$; (b) Same comparison for the maximum of
the ratio $r=\rD/\rmr$} \label{fig3}
\end{figure}

The analysis of the EOS is one of the most important issues
addressed in the present and future experiments. Recent combined
data\,\cite{WMAP3Y} suggest a value: $\omega_{\rm
exp}=-1.06^{+0.13}_{-0.08}\,.$ This result \textit{does} depend
on the assumption that the EOS parameter does \textit{not} evolve
with time or redshift, so it is not directly applicable to the
effective EOS of our model. Even so, we can find many scenarios
that are in good agreement with it, as shown in
Fig.{\ref{fig3}}(a). We see that the value of $\nu$ modulates the
behavior of the EOS, that can be QE-like (even though the $X$ is
phantom-like!, see (\ref{eEOS})), mimic that of a CC or present a
mild evolution from the phantom to the QE region.

All the curves in Fig.{\ref{fig3}}(a) satisfy (\ref{C1approx}),
presenting a stopping point and therefore (c.f.
Sect.\ref{sec:coin}) a maximum of $r(z)$. This ratio is plotted
in Fig.\ref{fig3}(b) in units of its current value ($r_0\approx
7/3$), showing that $r(z)$ remains bounded and $\sim r_0$ for
essentialy the entire Universe lifetime, which provides a natural
solution to the coincidence problem. Let us stress that these
features can occur even for $\nu=0$ (that is, for a strictly
constant $\CC$).

\section{Conclusions}

We have shown that the $\CC$XCDM model can be in good agreement
with present data and provide a solution to the cosmological
coincidence problem as well as a clear signature. In our opinion
the next generation of high precision cosmology experiments (DES,
SNAP, PLANCK)\,\cite{SNAP} should consider the possibility of a
composite DE with dynamics controlled by the running of the
cosmological parameters.
%%%%%%%%%%%%%%%%%%%%%%%%%%%%%%%%%%%%%%%%%%%%%%%%
%% BACKMATTER
%%%%%%%%%%%%%%%%%%%%%%%%%%%%%%%%%%%%%%%%%%%%%%%%
\begin{theacknowledgments}

This work has been supported in part by Ministerio de Eduaci\'on
y Ciencia of Spain (MEC) and FEDER under project
2004-04582-C02-01, and also by DURSI under 2005SGR00564. JG was
also supported by MEC under BES-2005-7803. The work of HS is
financed by the MEC and he thanks the Dep. ECM of the UB for the
hospitality.

\end{theacknowledgments}

%%%%%%%%%%%%%%%%%%%%%%%%%%%%%%%%%%%%%%%%%%%%%%%%%%%%%%%%%%%%%%%%%%%%%%%%%
\newcommand{\JHEP}[3]{{\sl J. of High Energy Physics } {JHEP} {#1} (#2)  {#3}}
\newcommand{\NPB}[3]{{\sl Nucl. Phys. } {\bf B#1} (#2)  {#3}}
\newcommand{\NPPS}[3]{{\sl Nucl. Phys. Proc. Supp. } {\bf #1} (#2)  {#3}}
\newcommand{\PRD}[3]{{\sl Phys. Rev. } {\bf D#1} (#2)   {#3}}
\newcommand{\PLB}[3]{{\sl Phys. Lett. } {\bf #1B} (#2)  {#3}}
\newcommand{\EPJ}[3]{{\sl Eur. Phys. J } {\bf C#1} (#2)  {#3}}
\newcommand{\PR}[3]{{\sl Phys. Rep } {\bf #1} (#2)  {#3}}
\newcommand{\RMP}[3]{{\sl Rev. Mod. Phys. } {\bf #1} (#2)  {#3}}
\newcommand{\IJMP}[3]{{\sl Int. J. of Mod. Phys. } {\bf #1} (#2)  {#3}}
\newcommand{\PRL}[3]{{\sl Phys. Rev. Lett. } {\bf #1} (#2) {#3}}
\newcommand{\ZFP}[3]{{\sl Zeitsch. f. Physik } {\bf C#1} (#2)  {#3}}
\newcommand{\MPLA}[3]{{\sl Mod. Phys. Lett. } {\bf A#1} (#2) {#3}}
%%%%%%%%%%%%%%%%%%%%%%%%%%%%%%%%%%%%%%%%%%%%%%%%%%%%%%%%%%%%%%%%%%%%%%%%%
%%%%%%%%%%%%%%%%%%%%%%%%%%%%%%%%%%%%%%%%%%%%%%%%%%%%%%%%%%%%%%%%%%%%%%%%%
\newcommand{\CQG}[3]{{\sl Class. Quant. Grav. } {\bf #1} (#2) {#3}}
\newcommand{\JCAP}[3]{{\sl JCAP} {\bf#1} (#2)  {#3}}
\newcommand{\APJ}[3]{{\sl Astrophys. J. } {\bf #1} (#2)  {#3}}
\newcommand{\AMJ}[3]{{\sl Astronom. J. } {\bf #1} (#2)  {#3}}
\newcommand{\APP}[3]{{\sl Astropart. Phys. } {\bf #1} (#2)  {#3}}
\newcommand{\AAP}[3]{{\sl Astron. Astrophys. } {\bf #1} (#2)  {#3}}
\newcommand{\MNRAS}[3]{{\sl Mon. Not.Roy. Astron. Soc.} {\bf #1} (#2)  {#3}}
%%%%%%%%%%%%%%%%%%%%%%%%%%%%%%%%%%%%%%%%%%%%%%%%%%%%%%%%%%%%%%%%%%%%%%%%%

%\endinput

\begin{thebibliography}{99}

\bibitem{Supernovae} R. A. Knop \textit{ et al.}, \APJ {598} {102} {2003};
A.G. Riess \textit{ et al.} \APJ {607} {2004} {665}.

\bibitem{WMAP3Y} D.N. Spergel \textit{et al.}, \textit{WMAP three year results: implications for
cosmology}, {astro-ph/0603449}.

\bibitem{LSS} M. Tegmark \textit{et al}, \PRD {69}{2004}{103501}.

\bibitem{Peebles84}  P.J.E. Peebles, \APJ {284}{1984}{439}.

\bibitem{weinRMP} S. Weinberg, \RMP {\bf 61} {1989}  {1}.

\bibitem{Copeland06} See e.g. E.J. Copeland, M. Sami, S. Tsujikawa, {hep-th/0603057},
and references therein.

\bibitem{Alam} U. Alam, V. Sahni, A.A. Starobinsky, \textit{JCAP} {0406} (2004)
{008}.

\bibitem{Jassal1} H.K. Jassal, J.S. Bagla, T. Padmanabhan, \PRD {72}{2005}{103503}.

\bibitem{SS12} J. Sol\`a, H. \v{S}tefan\v{c}i\'{c}, \MPLA {21} {2006}
{479}; \PLB {624}{2005}{147};  \textit{J. Phys.} {\bf A 39}
(2006) 6753; J. Sol\`a, \textit{J.Phys.Conf.Ser.} {\bf 39} (2006)
179.



\bibitem{GSS1} J. Grande, J. Sol\`a, H. \v{S}tefan\v{c}i\'{c}, \JCAP
{0608}{2006}{011}.


\bibitem{nova} I. Shapiro,  J. Sol\`{a}, \textit{JHEP} {\bf 0202}
{2002} {006}; \textit{Phys. Lett.} {\bf 475B} {2000} {236}.

\bibitem{PSW} R.D. Peccei, J. Sol\`{a}, C. Wetterich, \PLB {195} {1987}
{183}.


\bibitem{GSS2} J. Grande, J. Sol\`a, H. \v{S}tefan\v{c}i\'{c},
\textit{Composite dark energy: cosmon models with running
cosmological term and gravitational coupling}, {gr-qc/0609083}.

\bibitem{RGTypeIa}  I.L. Shapiro, J. Sol\`a, C. Espa\~na-Bonet, P. Ruiz-Lapuente,
\PLB {574} {2003} {149}; \textit{JCAP} {0402} (2004) {006}; I.L.
Shapiro, J. Sol\`a, \NPPS {127} {2004} {71}; astro-ph/0401015.

\bibitem{Ferreira97} P. G. Ferreira, M. Joyce, \PRD {58}{1998}{023503}.

\bibitem{SNAP} http://www.darkenergysurvey.org/; http://snap.lbl.gov/;
http://www.rssd.esa.int/index.php?project=Planck.

\end{thebibliography}
\end{document}